\documentclass[%
 reprint,
superscriptaddress,
showkeys,
 amsmath,amssymb,
 aps,
pre, 
]{revtex4-2}

\usepackage{graphicx}% Include figure files
\usepackage[caption=false]{subfig} %?

\usepackage{dcolumn}% Align table columns on decimal point
\usepackage{bm}% bold math
\newcommand{\ee}{\end{equation}} 
\newcommand{\be}{\begin{equation}}

\usepackage[mathscr]{euscript}  
\usepackage{xcolor}  

\usepackage{tcolorbox}
\usepackage{comment}
\usepackage{float}

\makeatletter
\newsavebox{\@brx}
\newcommand{\llangle}[1][]{\savebox{\@brx}{\(\m@th{#1\langle}\)}%
  \mathopen{\copy\@brx\kern-0.5\wd\@brx\usebox{\@brx}}}
\newcommand{\rrangle}[1][]{\savebox{\@brx}{\(\m@th{#1\rangle}\)}%
  \mathclose{\copy\@brx\kern-0.5\wd\@brx\usebox{\@brx}}}
\makeatother

\begin{document}

\preprint{ApS/123-QED}

\title{Steady-state moments under resetting to a distribution}

\author{Kristian St\o{}levik Olsen}

\affiliation{Nordita, Royal Institute of Technology and Stockholm University, Hannes Alfvéns väg 12, 23, SE-106 91 Stockholm, Sweden\\}

\affiliation{Institut für Theoretische Physik II - Weiche Materie, Heinrich-Heine-Universität Düsseldorf, D-40225 Düsseldorf, Germany}

\begin{abstract}
The non-equilibrium steady states emerging from stochastic resetting to
a distribution is studied. We show that for a range of processes, the steady-state moments can be expressed as a linear combination of the moments of the distribution of resetting positions. The coefficients of this series are universal in the sense that they do not depend on the resetting distribution, only underlying dynamics. We consider the case of a Brownian particle and a run-and-tumble particle confined in a harmonic potential, where we derive explicit closed-form expressions for all moments for any resetting distribution. Numerical simulations are used to verify the results, showing excellent agreement. 
\end{abstract}

\pacs{Valid pACS appear here} % pACS, the physics and Astronomy  Classification Scheme.
\maketitle
%%%%%%%%%%%%%%%%%%%%%%%%%%

\section{Introduction}

Large fluctuations are present in almost all systems in Nature, and are typically unavoidable in small systems. Instantaneous and large jumps in a systems state can have drastic implications for the systems dynamics, and potentially drive it out of thermal equilibrium. Stochastic resetting is one example of large and sudden fluctuations, whereby a systems state is typically brought back to its initial state at a constant rate \cite{evans2011diffusion,evans2020stochastic}. The study of (non-equilibrium) steady states emerging from stochastic resetting has gained significant attention in the past decade due to both its broad relevance in diverse scientific disciplines, and the analytically tractable steady states that emerge. Most famous is perhaps the application of resetting in search processes, where it has been shown that resets, when optimized, can expedite a search process \cite{evans2011diffusion, reuveni2016optimal,pal2017first, pal2019first, ahmad2019first, tal2020experimental}. In biology resetting can be found across multiple scales, ranging from protein-bound search processes \cite{cui2019facilitated} to the migratory and foraging patterns of animals \cite{grecian2018understanding,pal2020search}. In physics, fundamental aspects have been in regards to the non-equilibrium nature of resetting, such as relaxation dynamics to non-equilibrium steady states \cite{majumdar2015dynamical,singh2020resetting}, and stochastic thermodynamics \cite{fuchs2016stochastic,gupta2020work,gupta2022work,pal2017integral,pal2021thermodynamic, prr,sunil2023cost,goerlich2023experimental}.

The paradigmatic example of stochastic resetting in physics is that of a reset Brownian motion \cite{evans2011diffusion,evans2011optimal,evans2020stochastic,sherman1958limiting,feller1954diffusion}. Since the first studies of resetting in its modern form more than a decade ago, a myriad of generalizations have surfaced, including Brownian motion in potentials \cite{pal2015diffusion,ray2020diffusion}, processes where the diffusivity switches between multiple possible values \cite{bressloff2020switching}, fractional Brownian processes \cite{majumdar2018spectral,wang2021time}, and in active matter models like active Brownian and run-and-tumble particles \cite{evans2018run,kumar2020active, tucci2022first,santra2020run}. Steady states have also been studied under a variety of resetting schemes, for example time-dependent resetting rates \cite{pal2016diffusion}, non-instantaneous resets \cite{radice2022diffusion}, non-Poissonian waiting times \cite{eule2016non, nagar2016diffusion,radice2022diffusion}, or resetting mediated by an external trap \cite{gupta2020stochastic} to mention a few. For a recent review of stochastic resetting and its applications, see \cite{evans2020stochastic}.

Another resetting scheme of high degree of relevance to both theory and experiments is one where the resetting position is random and drawn from a distribution $p_R(x)$.  Recently, it has been argued that only for resetting distributions that are not delta-peaked does the standard path-based framework of stochastic thermodynamics make sense for resetting systems \cite{prr}. It is also highly natural to assume that resets cannot be performed with perfect precision, neither in experiments or in natural systems. In addition, from a theoretical perspective, the resetting distribution $p_R(x)$ introduces new lengthscales into the problem, which can give rise to interesting phenomena.

\begin{figure}
    \centering
    \includegraphics[width = 7.2cm]{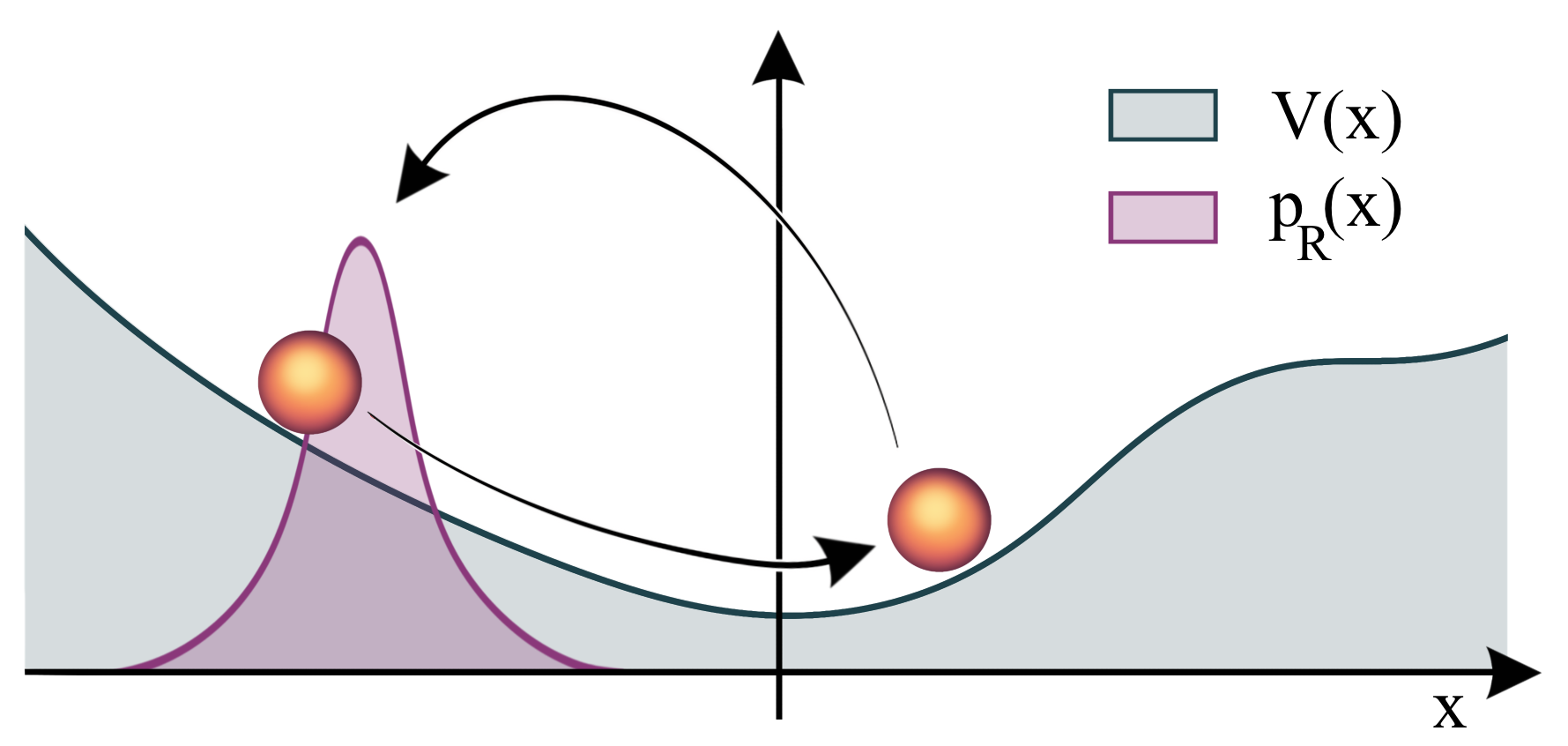}
    \caption{Sketch of the system under consideration. A particle moving in a potential $V(x)$ undergoes resetting with constant rate $r$ to a resetting distribution $p_R(x)$. Multiple lengthscales are present in the problem, competing to produce interesting steady-state properties.}
    \label{fig:sketch}
\end{figure}

In this paper we characterize the steady-states of resetting processes where the reset position is drawn from a distribution $p_R(x_0)$. See Fig. (\ref{fig:sketch}) for a sketch of the system considered. We consider Poissonian resets, where the waiting time between resets is distributed exponentially with constant rate $r$. The exact steady-state distribution can in general be rather arduous to compute for any interesting choice of $p_R(x_0)$, and using it to calculate moments and cumulants can be tricky from a practical standpoint. Here we provide an alternate approach, where we find exact closed-form formulas for all the moments in the steady state, assuming that the moments of the resetting distribution $\langle x_0^n\rangle_R$ are known \emph{a priori}. The main results take the form
\begin{equation}\label{eq:genform}
    \langle x^n\rangle_* = \sum_{j=0}^n C_{nj} \langle x_0^j\rangle_R
\end{equation}
where the asterisk denotes steady states and where the coefficients $\{C_{nj}\}_{j=1}^n$ depend only on resetting rate and underlying system parameters. The fact that the coefficients will not depend on the shape of the resetting distribution makes the results rather universal, and can be used for gain insight into the steady states under any choice of resetting distribution once the coefficients $C_{nj}$ are known. The presented results could be verified experimentally using optical tweezers, and be of relevance to further theoretical studies of distributed resetting for example in thermodynamical settings \cite{prr}.

This paper is organized as follows. Section \ref{sec:renewal} introduces the (last) renewal equation, and uses it to prove the main result of the paper, Eq. (\ref{eq:genform}), for the steady state moments for a class of processes.  Section \ref{sec:PB} derives the series coefficients $C_{nj}$ for Brownian motion in a harmonic potential, i.e. for an Ornstein-Uhlenbeck process.  In section \ref{sec:RTP} we consider a run-and-tumble particles in a harmonic potential, which may be though of as a non-Markovian version of the Ornstein-Uhlenbeck process, driven by a telegraphic noise.  Section \ref{sec:concl} offers concluding remarks.

\section{The renewal equation approach}\label{sec:renewal}

The standard route to find the propagator $p(x,t|x_0)$ in the presence of resetting is through the (last) renewal equation \cite{evans2020stochastic}
\begin{align}
    p(x,t|x_0) = & e^{- r t} p_0(x,t|x_0) \nonumber\\
    &+ r \int_0^t  d \tau e^{- r \tau} \int dx_0 p_R(x_0) p_0(x,\tau|x_0).
\end{align}
Here $p_0(x,t)$ is the propagator of the underlying system in absence of resetting ($r=0$). The first term corresponds to trajectories where no resetting took place (which happens with probability $e^{- rt}$). The second term takes into account trajectories where resetting does happen, and makes use of the fact that at every reset all dynamical memory is erased. Therefore one only needs information regarding the last reset, which took place at time $t-\tau$, before evolving freely for an exponentially distributed waiting time $\tau$ from a reset position $x_0$ drawn from $p_R(x_0)$.

In the late-time regime, the renewal equation predicts the following steady state
\begin{align} \label{eq:ss}
        p_{*} (x) &= r \int_0^\infty  d \tau e^{- r \tau} \int dx_0 p_R(x_0) p_0(x,\tau|x_0)\\
        &= r \int dx_0 p_R(x_0) \tilde{p}_0(x,r|x_0)
\end{align}
where $\tilde{p}_0(x,s|x_0)$ is the Laplace transform of $p_0$. For many interesting cases beyond free Brownian motion, performing the above integrals analytically at best arduous and in the worst case impossible in closed form. Even for simple case of freely diffusive Brownian particles with Gaussian resetting distribution, the steady state density becomes rather involved \cite{prr}. Here we propose a method based on moments that is useful when the underlying process is more involved, such as in the case of distributed resets for diffusion in a potential.

The expression in Eq. (\ref{eq:genform}) can be shown to follow from the renewal equation for a certain class of processes. From the last renewal equation we know that the steady-state moments can  in general be written as 
\begin{align}
    \langle x^n\rangle_* &= r\int_0^\infty d\tau e^{- r \tau} \int dx_0 \int dx x^n p_0(x,\tau|x_0) p_R(x_0)
\end{align}
To proceed, we consider a class of processes where the propagator satisfies $ p_0(x,\tau|x_0) =  p_0(x- x_0 g(\tau),\tau)$, for some time-dependent function $g(\tau)$ that may be specific to each process. For spatially homogeneous processes, such as free Brownian motion, Levy flights etc, we have $ p_0(x,\tau|x_0) =  p_0(x-x_0,\tau)$ and hence $g(\tau) = 1$. For the Ornstein-Uhlenbeck process with relaxation-time $\tau_\text{rel}$, $g(\tau) = \exp(- \tau/\tau_\text{rel})$. Making a change of coordinates to $y = x-x_0 g(\tau)$ we find 
\begin{equation}\label{eq:tempmom}
    \langle x^n\rangle_* =  r \int d\tau e^{-r \tau}\int  dx_0\int dy [y+x_0 g(\tau)]^n  p_0(y,\tau) p_R(x_0) 
\end{equation} 
Using the binomial formula we can write 
\begin{equation} \label{eq:bino}
 [y+x_0 g(\tau)]^n  = \sum_{j=0}^n {n \choose j}  y^{n-j} g^j(\tau) x_0^j
\end{equation} 
Combining Eq. (\ref{eq:tempmom}) and Eq. (\ref{eq:bino}) one immediately finds Eq. (\ref{eq:genform}), where the expansion coefficients take the form
\begin{equation} \label{eq:coefs}
    C_{n,j} = r{n\choose j}\int dy  \int_{0}^\infty d\tau e^{- r\tau} y^{n-j}  g^j(\tau) p_0(y,\tau) 
\end{equation}
We see that the coefficients depends on the resetting rate, and in addition on the propagator of the underlying process through $p_0(y,\tau)$ and $g(\tau)$. In particular, they are related to the Laplace transform of the moments of the underlying process without resetting. Importantly, the coefficients do not depend on the resetting distribution, and therefore holds universally for any choice of $p_R(x)$.

\begin{figure*}
    \centering
    \includegraphics[width = 12cm]{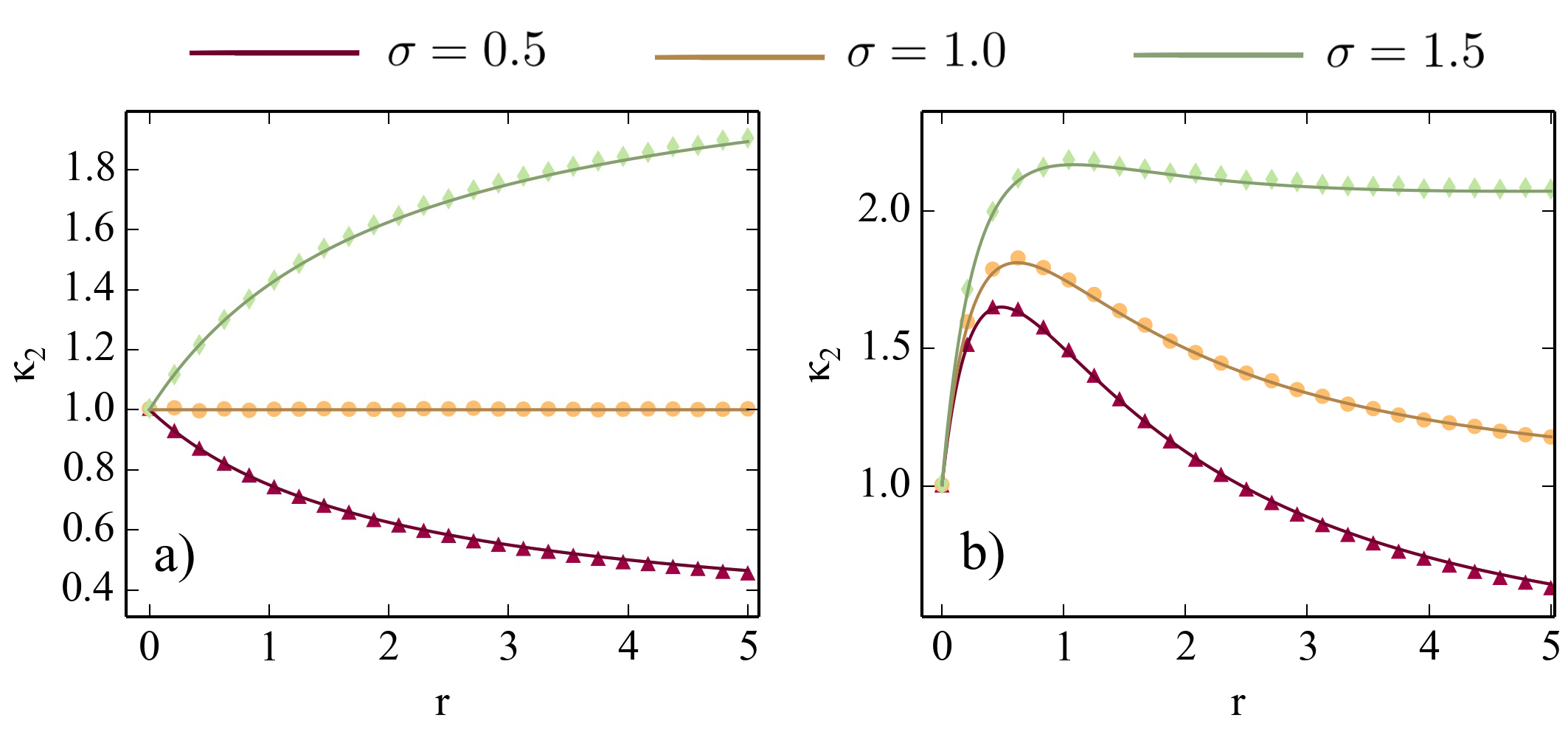}
    \caption{Steady state variance for a Brownian particle in a harmonic potential, for Gaussian resetting distribution with mean $z$ and variance $\sigma^2$. Panel a) shows resetting centered at the origin $z =0$, while panel b) shows the case $z = 3$. Dots show numerical simulations, while solid curves correspond to Eq. (\ref{eq:pbss_var}). Parameters are set to $k = \mu = D = 1$.  }
    \label{fig:gauss_k2}
\end{figure*}
\subsection{Spatially homogeneous processes}
Before we move on to study cases where the above approach is useful, we briefly mention some properties of spatially homogeneous systems. Homogeneity sets $g(t) = 1$, which will in some cases de-couple the space- and time-integration in Eq. (\ref{eq:coefs}). For example, we have the following identities
\begin{align}
    C_{nn} &= 1 \label{eq:id1}\\
    C_{n-1,n-2} &= \frac{n-1}{n}C_{n,n-1}\label{eq:id2}
\end{align}
which follow immediately from Eq. (\ref{eq:coefs}). The latter identity is a consequence of the fact that for homogeneous systems the integrand in Eq. (\ref{eq:coefs}) depends on $n$ and $j$ only through the combination $n-j$. This gives rise to the more general identity
\begin{equation}
    {n\choose n-k} C_{n
    -\ell,n-\ell-k} = {n-\ell\choose n-\ell-k} C_{n,n-k}\label{eq:id3}
\end{equation}
These properties of the series coefficients imply that the cumulants in the steady state are linearly related to the corresponding cumulants of the resetting distribution. For example, for the variance $\kappa_2$ and the third cumulant $\kappa_3$, we have 
\begin{align}
    \kappa_2 &= \kappa_2^R  + C_{2,0} - \frac{C_{2,1}^2}{4}  \label{eq:homcum2}\\
    \kappa_3 &=\kappa_3^R + C_{3,0}  - 3 C_{2,0}C_{1,0} +2 C_{1,0}^2 \label{eq:homcum3}
\end{align}
Hence the variance and the skew, as measured by $\kappa_3$, are directly "inherited" from $p_R(x)$, plus terms that comes from the dynamics in between resetting events. Another way to look at this is that the cumulants (except the mean) are simply additive combinations of the corresponding cumulant of the resetting distriution and the value one would obtain if the resetting was to a fixed position $p_R(x) = \delta(x-x_r)$. This is because in homogeneous systems exactly where the particle is reset does not matter for cumulants of higher order than $1$ due to spatial translational invariance. In this sense, the case of spatially homogeneous processes are not all that interesting to study under the effect of a distribution of resetting positions, as most properties are transferred directly from $p_R(x)$ or the corresponding resetting process when $p_R(x) = \delta(x-x_r)$.

However, if the system is \emph{not} homogeneous there are bound to be some lengthscales already present in the system. These can interfere with the scales introduced by the resetting distribution, giving rise to more complex steady-state properties. For more general processes, such as Brownian motion in a harmonic potential, $g(t) \neq 1$ and the coefficients do not satisfy the above relations Eqs. (\ref{eq:id1}, \ref{eq:id2}, \ref{eq:id3}).

While Eq. (\ref{eq:genform}) and (\ref{eq:coefs}) shows the general form of the steady-state moments, these equations are not always convenient as a practical way to derive $C_{n,j}$ for specific cases, as the integrals involved in Eq. (\ref{eq:coefs}) are not always trivial. The approach we will take, relying on the knowledge that Eq. (\ref{eq:genform}) holds, will be to identify the coefficients $C_{nj}$ rather easily directly from the master equation for a range of processes.

\section{Distributed resets in a harmonic potential}\label{sec:PB}

In this section we consider Brownian motion in a harmonic potential with resetting to a distribution $p_R(x)$. During a timestep $dt$, the particle has a probability $r dt$ to follow the underlying dynamics
\begin{equation}\label{eq:eom}
x(t+dt) =x(t) + \mu F(x)  dt + \sqrt{2D}\Delta W(t),
\end{equation}
and a complimentary probability $1-r dt$ to be reset 
\begin{equation}
x(t+dt) = x_r.
\end{equation}
Here $x_r$ is the resetting position which we draw from a resetting distribution $p_R(x)$ at every resetting event. In Eq. (\ref{eq:eom}), $\Delta W(t)$ is the increment of a Wiener process, which has mean zero and variance $\langle \Delta W(t)^2 \rangle  =\sqrt{dt}$. $F(x) = - V'(x)$ is a conservative force field originating from a potential $V(x)$, $\mu = 1/\gamma$ is the mobility (inverse friction coefficient), and $D$ is the diffusivity.

The associated master equation for the density $p(x,t| x_0,t)$ can be obtained from noting that in a small step $dt$ there are two ways the density can be updated \cite{evans2020stochastic}
\begin{equation}
    p(x,t+dt|x_0,0) = (1-r dt)\left \langle  p(x - dx, t | x_0)\right\rangle_{dx} + r dt p_R(x)
\end{equation}
Here the first term takes into account trajectories where the particle at time $t$ was located at position $x' = x - dx$ before taking a (random) step $dx$ without resetting. This happens with probability $1-r dt$. With probability $r dt$, the particle resets to a position drawn from $p_R(x)$. 
Expanding to first order in $dt$, meaning second order in $dx$ due to the properties of the Wiener increments, one finds in the $dt \to 0$ limit the master equation 
\begin{align} \label{eq:master}
    \partial_t p(x,t) =& \partial_x \left[ -\mu F(x) p(x,t) + D \partial_xp(x,t) \right] \nonumber \\
    &- r p(x,t) + r p_R(x)
\end{align}
where we have written $p(x,t|x_0,t_0) = p(x,t)$ for simplicity.

The time-evolution of any generic observable $\langle f(x) \rangle$  can be derived using this master equation. If the time evolution operator in the absence of resetting is denoted $\hat{\mathcal{L}} $, one generally has 
\begin{equation}
    \partial_t \langle f(x) \rangle =  \int dx  f(x) \hat{\mathcal{L}} p(x,t) - r \langle f(x) \rangle + r \langle f(x) \rangle_R
\end{equation}
where the subscript $R$ denoted averages calculated using the resetting distribution $p_R(x)$. If a steady state can be reached at late times $\partial_t \langle f(x) \rangle_*  = 0$, one finds
\begin{equation}\label{eq:obs}
    \langle f(x) \rangle_* =  \langle f(x) \rangle_R +  r^{-1}   \langle\hat{\mathcal{L}} ^\dagger f(x)  \rangle_*
\end{equation}
where the asterisk denotes steady states.  
For monomial functions $f(x) = x ^n$ Eq. (\ref{eq:obs}) together with Eq. (\ref{eq:master}) results in the following hierarchy for the steady-state moments
\begin{equation}\label{eq:hier}
        \langle x^n \rangle_* =  \langle x^n \rangle_R  -\frac{ n \mu}{r} \langle x^{n-1} \partial_x V(x)\rangle_* + \frac{n (n-1) D}{r}  \langle x^{n-2} \rangle_*
\end{equation}
For a harmonic potential $V(x) = \frac{1}{2}k x^2$ the hierarchy simplifies, becoming
\begin{equation}
        \langle x^{n} \rangle_*  =\frac{ \langle x^{n} \rangle_R}{1+\frac{ n \mu k}{r}}  + \frac{n (n-1) D}{r(1+\frac{ n \mu k}{r})}  \langle x^{n-2} \rangle_* 
\end{equation}
Rather than calculating the integrals in Eq. (\ref{eq:coefs}), we can solve this second-order recursive equation directly using standard methods, resulting in
\begin{widetext}
\begin{equation}\label{eq:brownianmom}
     \langle x^{n} \rangle_*  = \sum_{j \in 2 \mathbf{N}}^n\left(\prod_{i\in 2 \mathbf{N}, i < j}\frac{(n-i) (n-i -1) }{1+\frac{ (n-i) \mu k}{r}}\right) \left( \frac{D}{r}\right)^{j/2} \frac{ 1}{1+\frac{ ({n-j}) \mu k}{r}} \:  \langle x^{n-j} \rangle_R
\end{equation}
\end{widetext}

where the sum is over positive even numbers, $ 2\mathbf{N}$,(with zero included) up to and including $n$, and the product is over positive even numbers up to the largest even number smaller than $j$. Note that the order of summation is opposite to that of Eq. (\ref{eq:genform}), which can easily be reversed by summing over  $i = n-j$ in stead. This gives the expansion coefficients for Brownian motion in a harmonic potential, valid for any resetting distribution $p_R(x)$.

From Eq. (\ref{eq:brownianmom}) any moment or cumulant can be obtained directly. For example, the mean takes the simple form
\begin{equation}
      \langle x \rangle_*   = \frac{ \langle x \rangle_R}{1 + \mu k /r}
\end{equation}
 where we see that due to the potential the steady-state mean is always less than the mean of the resetting distribution. The variance takes the form
 \begin{equation}\label{eq:pbss_var}
     \kappa_2 = \frac{2 D / r +  \langle x^2 \rangle_R}{1 + 2\mu k /r} -\frac{ \langle x \rangle_R^2}{(1 + \mu k /r)^2}
 \end{equation}
This variance can display non-monotonic behavior as a function of resetting rate $r$ resetting with mean $\langle x\rangle_R$ non-zero. As a concrete example, consider resetting to a Gaussian distribution $p_R(x) = \mathcal{N}(x,\sigma)$ with mean $z$ and variance $\sigma^2$.  Fig. (\ref{fig:gauss_k2} a) shows the variance as a function of resetting rate for centered resetting $z=0$, while (\ref{fig:gauss_k2} b) shows the same quantity for $z\neq 0$. While resetting to a position with mean centered at the potential minimum gives rise to monotonic increase of decrease in the variance, a non-centered resetting $z\neq 0$ gives rise to non-monotonic variance with a global maximum at a critical resetting rate. High values of the resetting width $\sigma$, this non-monotonic behavior can be erased.  

The skew of the steady-state also shows interesting behaviors. Using again Eq. (\ref{eq:brownianmom}) for the third moments and combining them into the third cumulant, we find

\begin{align}\label{eq:pbss_kappa}
        \kappa_3  = & \frac{2 r^3 \langle x \rangle_R^3}{(k \mu +r)^3}+\frac{6 D
   \langle x \rangle_R r}{(k \mu +r) (3 k \mu +r)}
   \nonumber\\
   & -\frac{3 \langle x \rangle_R r (2 D+\langle x^2 \rangle_R r)}{(k \mu
   +r) (2 k \mu +r)}
   +\frac{r}{3 k \mu +r}\langle x^3 \rangle_R 
\end{align}

\begin{figure}
    \centering
    \includegraphics[width = 8.4cm]{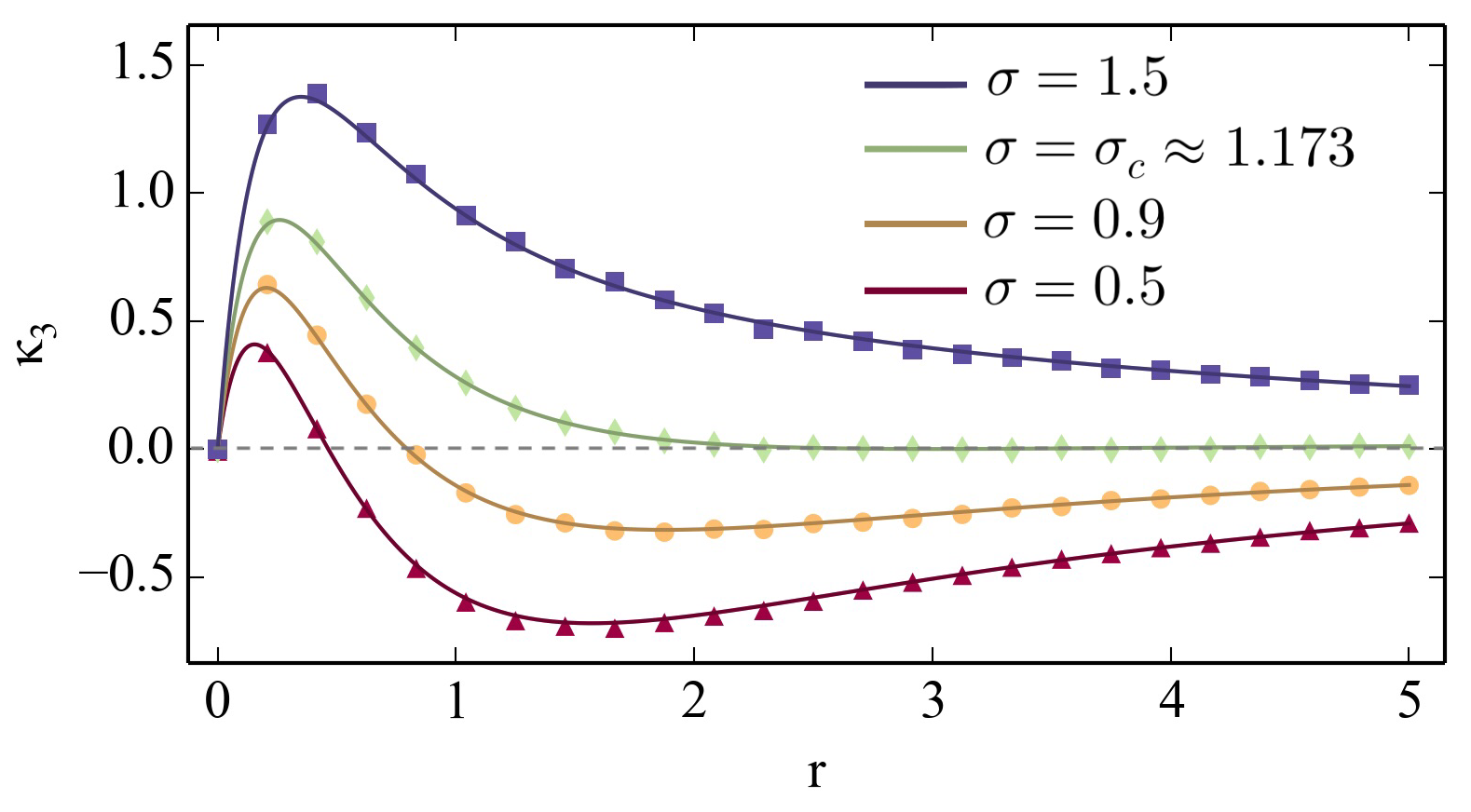}
    \caption{Steady state third cumulant from Eq.  (\ref{eq:pbss_kappa}) for an Ornstein-Uhlenbeck process with Gaussian resetting $p_R \sim\mathcal{N}(z,\sigma)$, where parameters used are $k = \mu = D = 1$. For sharp resetting to a point high in the potential landscape, $z = 3$ in this case, the skew of the distribution may change sign at a critical resetting rate given by Eq. (\ref{eq:rc}). According to Eq. (\ref{eq:k3ineq}), this will only happen for $\sigma < \sigma_c \approx 1.173$. Dots show numerical simulations, while solid curves correspond to Eq. (\ref{eq:pbss_kappa}).   }
    \label{fig:gauss_k3}
\end{figure}

Just like the variance, the skew may also display non-monotonic behavior. More interestingly,  in some parameter regions the skew even changes sign as a function of resetting rate; for $r<r_c$ the skew is positive, while for $r > r_c$ the skew is negative. Returning again to the case of Gaussian resetting $p_R \sim \mathcal{N}(z,\sigma)$ one can find the critical resetting rate $r_c$ where the skew is zero to be
\begin{align}\label{eq:rc}
    r_c =& \frac{\sqrt{k^3 \mu ^3 z^2 \left(24 D+k \mu  \left(z^2-24 \sigma ^2\right)\right)}}{6 \left(D-k \mu  \sigma ^2\right)} \nonumber \\
    & - \frac{k \mu  \left(6
   D+k \mu  \left(z^2-6 \sigma ^2\right)\right)}{6 \left(D-k \mu  \sigma ^2\right)}
\end{align}
This zero only exists when the resetting width is sufficiently small, namely
\begin{equation}\label{eq:k3ineq}
     \sigma^2 <  \sigma_c^2 \equiv   \frac{D }{k \mu} + \frac{z^2  }{24} 
\end{equation}
where it is assumed that $z \neq 0$. For larger resetting widths $\sigma > \sigma_c$ the third cumulant may still be non-monotonic, as seen in Fig. (\ref{fig:gauss_k3} c), but will not change sign as resetting rate is varied.

\section{Resetting of run-and-tumble particles}\label{sec:RTP}

This section derives the series coefficients entering into Eq. (\ref{eq:genform}) for an active run-and-tumble particle in a harmonic potential. In active particle systems non-trivial steady states emerge when the particles are put into confinement, either due to interactions with solid boundaries or smooth potentials \cite{elgeti2015run, gompper20202020}. Intriguing phenomena has been studied in this context in the past, ranging from accumulation and trapping of individual particles at confining boundaries \cite{moen2022trapping,bressloff2023trapping,debnath2021escape,souzy2022microbial,volpe2014simulation,yang2014aggregation,caprini2018active,caprini2019active}, to collective phenomena only possible due to the presence of confinement \cite{olsen2022collective, kumar2019trapping, olsen2020escape}. Understanding the interplay between the activity of the particles and the confinement is crucial for elucidating the complex behavior of active matter in realistic environments. On top of confinement, we here include the effects of distributed stochastic resetting. In addition to a model of active matter, the  model considered here may also be interpreted as distributed resetting in a non-Markovian Ornstein-Uhlenbeck process driven by a telegraphic rather than Gaussian white noise.

\subsection{Stochastic equation and coupled master equations}
We consider a RTP following the stochastic dynamics 
\begin{equation}\label{eq:lang_r}
    x_{t+dt} = \left\{
	\begin{array}{ll}
		 x_t + f(x_t) dt + v_0 \kappa(t) dt  & \textnormal{with probability } 1-r dt \\
		x_r  &\textnormal{with probability } r dt
	\end{array}
\right.
\end{equation}
where $ \kappa(t)$ is a telegraph process that switches between values $+1$ and $-1$ with a constant rate $\alpha$. Hence, $\kappa$ obeys
\begin{equation}
     \kappa(t+dt)  = \left\{
	\begin{array}{ll}
		 \kappa(t)  & \textnormal{with probability } 1-\alpha dt \\
		-\kappa(t)   &\textnormal{with probability } \alpha dt
	\end{array}
\right.
\end{equation}
The value of $\kappa(t)$ indicates the particles direction of motion in which it swims with speed $v_0$.  In the above, $f(x_t)$ a time-independent external force which we will assume to be conservative $f(x) = - \partial_x V(x)$ for some potential $V(x)$. The resetting position $x_r$ is as before drawn at each reset from the resetting distribution $p_R(x)$.

Upon resetting also the internal velocity state $\kappa(t)$ of the RTP is also subject to resetting, and we will assume that motion in the positive (negative) direction is chosen with probability $\rho_+$ (resp. $\rho_-$) at each reset. Hence, in addition to the underlying telegraphic process where $ \kappa(t) $ changes sign with probability $\alpha dt$ in a timestep of size $dt$, the process is at each resetting updated according to 
\begin{equation}\label{eq:lang_r}
    \kappa(t+dt) = \left\{
	\begin{array}{ll}
		 +1 & \textnormal{with probability } \rho_+ \\
		 -1  &\textnormal{with probability } \rho_-
	\end{array}
\right.
\end{equation}
We assume that the process is also initialized according to  the probabilities $\rho_\pm$.  The state of the system is specified by the probability density $p_z(x)$, where $z = \pm 1$ corresponds to the two swimming directions. The master equation for this process can be conveniently expressed as
\begin{align}
\partial_t p_+ = - v \partial_x p_+ - \mu \partial_x [fp_+]   - (\alpha + r) p_+ + \alpha p_- + r p_R \rho_+  \\
\partial_t p_- =  v \partial_x p_- - \mu \partial_x [fp_-] - (\alpha + r) p_- + \alpha p_+ + r p_R \rho_-
\end{align}
In the case of a confining potential, it is known that the densities $p_\pm$ separately become stationary \cite{dhar2019run}. Since resetting effectively confines the process by curbing large excursions, we will assume that $\partial_t p_\pm = 0$ in the steady-state, even if $V(x) = 0$. In this model, a non-equilibrium steady state is expected to be reached for two reasons; one is the resetting dynamics which produces steady states even in passive systems. In addition there is the effect of activity, which further pushes the system away from thermodynamic equilibrium.

\begin{figure*}
    \centering
    \includegraphics[width = 15cm]{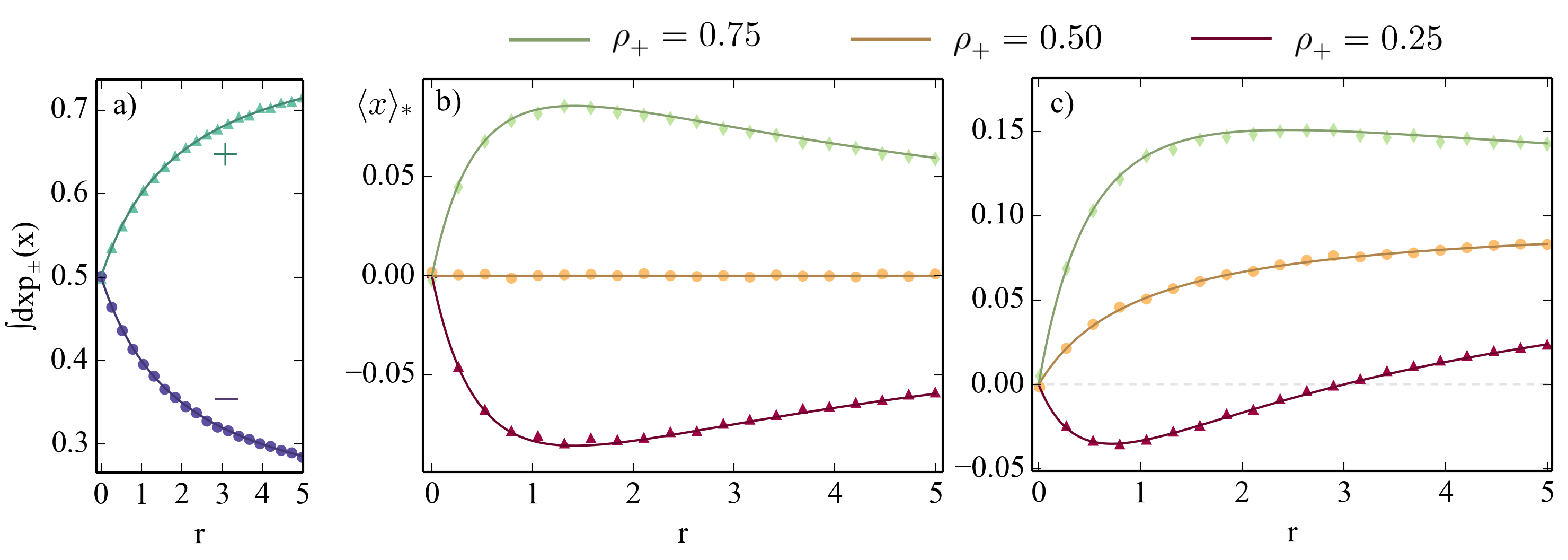}
    \caption{Comparisons between numerical simulations and theory for harmonically confined run-and-tumble with resetting to a Gaussian distribution $p_R(x) = \mathcal{N}(z,\sigma)$. a) Probability of finding the particle moving to the right $(+)$ or to the left $(-)$ in the steady state, given by Eq. (\ref{eq:plmin}), for $\rho_+ = 0.8$. b) Steady-state mean $\langle x \rangle _*$ as a function of resetting rate for $z = 0$. We see that the mean is generally not monotonic for asymmetric velocity resetting $\rho_\pm \neq 0$. c)   Steady-state mean $\langle x \rangle_*$ as a function of resetting rate for $z = 3$. Points correspond to numerical simulations, while solid lines theory. Parameters used are $\alpha = \mu = k = v_0 = 1$.   }
    \label{fig:rtp_k1}
\end{figure*}

\subsection{Steady-state moments}\label{sec:hierRTP}

In order to obtain moments for the RTP case, we proceed similarly to the passive Brownian case. We define
\begin{align}
    m_n &= \sum_{z= \pm} \int dx x^n p_z = \int dx x^n (p_+ + p_-) \label{eq:mns}\\
    d_n &= \sum_{z= \pm} \int dx x^n z p_z = \int dx x^n (p_+ - p_-) \label{eq:dns}
\end{align}
where $\{m_n\}$ are the full bare moments of the spatial density $p = p_+ + p_-$, while $\{d_n\}$ measures the difference in the $n$'th moment between the right- and left-moving density. To emphasize the notation of this section, Eq. (\ref{eq:genform}) will now take the form
\begin{equation}
   m_n = \sum_{j=0}^n C_{nj} m^R_j
\end{equation}
where $m_1 = \langle x \rangle_*$, $m_2 = \langle x^2 \rangle_*$ \emph{et cetera}. While $d_n$ does not have as clear of an interpretation as the bare moments $m_n$, it does contain information about the steady state. For example, $d_0$ contains information regarding the mean current in the system, i.e. gives the probability of finding the particle in the right- or left-moving states in the steady state through $\int dx p_\pm (x) = \frac{1\pm d_0}{2}$, where we also used that by normalization $m_0 = 1$.

For a harmonic potential $V(x) = \frac{1}{2} kx^2$ the master equation gives rise to the following steady-state hierarchy for the moments
 \begin{align}
    m_n &= \frac{r }{r + n k\mu }m_n^R  + \frac{ n v_0 }{r + n k\mu  }d_{n-1}     \\
    d_n &= \frac{r (\rho_+ - \rho_-) }{2 \alpha + r+ n k\mu}m_n^R  + \frac{ n v_0 }{2 \alpha + r+ n k\mu}m_{n-1}
 \end{align}
 Here we immediately see that $ d_0 = (r (\rho_+ - \rho_-) )/(2 \alpha + r)$, giving the probability of finding the particle swimming to the right (+) or left(-) as
 \begin{equation}\label{eq:plmin}
     \int dx p_\pm(x) = \frac{2 \alpha + r \pm r (\rho_+ - \rho_-)}{4 \alpha + 2r }
 \end{equation}

This is shown in Fig. (\ref{fig:rtp_k1} a). As $d_n$ in Eq. (\ref{eq:dns}) is given entirely in terms of the bare moments $m_{n-1}$, we can easily find the closed hierarchy 
 \begin{align}
     m_n &= \frac{r m_n^R }{ r + n k \mu}  +  \frac{ n v_0 }{ r + n k \mu} \frac{r (\rho_+ - \rho_-)  }{2\alpha  + r+ (n-1) k\mu }  m_{n-1}^R \\
    &+ \frac{ n (n-1) v_0^2 }{ ( r + n k \mu)(2 \alpha  + r+ (n-1) k\mu ) }   m_{n-2}  \nonumber
 \end{align}
Although this recursive relation has somewhat complex coefficients, it is of the general form $ m_n =  b_n + g_n m_{n-2}$ just as in the passive case, where we defined coefficients
\begin{align}
    b_n &= \frac{r m_n^R }{ r + n k \mu}  +  \frac{ n v_0 }{ r + n k \mu} \frac{r (\rho_+ - \rho_-)  }{2\alpha  + r+ (n-1) k\mu }  m_{n-1}^R \\
    g_n & =  \frac{ n (n-1) v_0^2 }{ ( r + n k \mu)(2 \alpha  + r+ (n-1) k\mu ) }
\end{align}
Here $m_n^R = \langle x^n\rangle_R$ is again the $n$'th moment of the resetting distribution.  In this notation, the solution reads
\begin{equation}
    m_n = \sum_{j \in 2 \mathbf{N}}^n\left(\prod_{i\in 2 \mathbf{N}, i < j} g_{n-i}\right) b_{n-j}.
\end{equation}

\begin{figure*}
    \centering
    \includegraphics[width = 14.5cm]{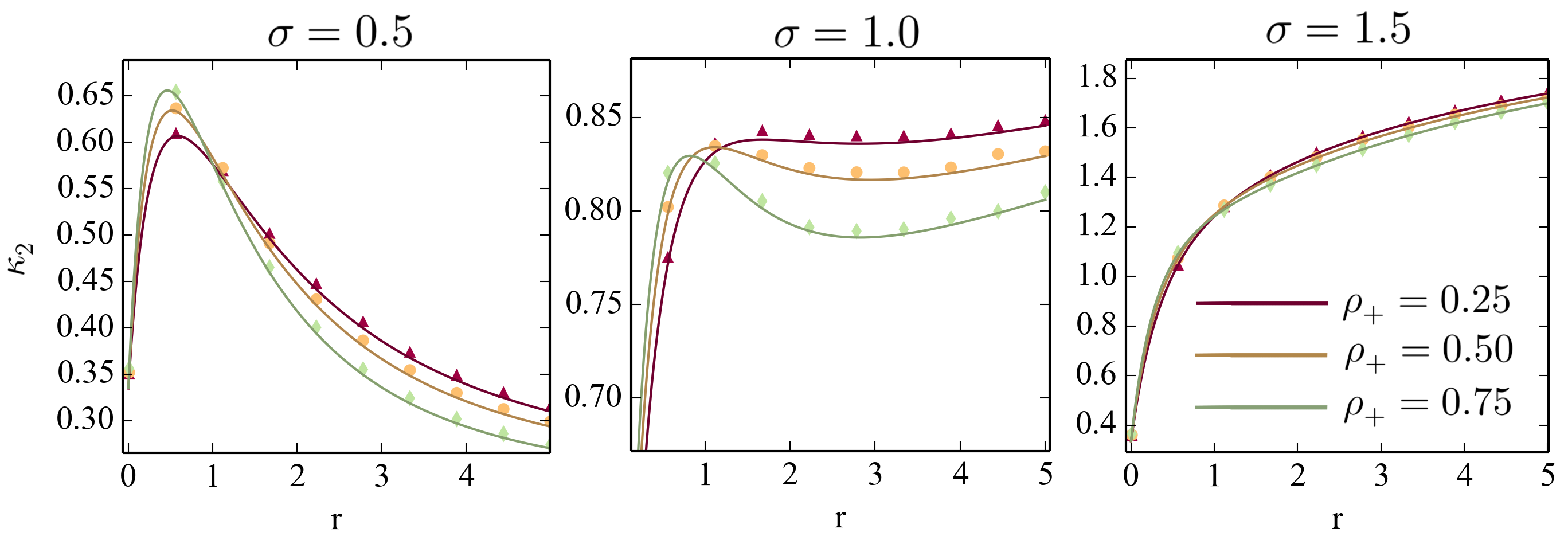}
    \caption{Steady state variance of a run-and-tumble particle under resetting to a Gaussian with mean $z = 2$ and variance $\sigma^2$,  in the presence of a harmonic potential.  Different colored curves correspond to different asymmetries in the internal velocity states, controlled by $\rho_+$.  For high resetting width $\sigma$, the variance becomes monotonically growing, with little distinction between different values of $\rho_+$. Parameters are set to $k = \mu = \alpha = v_0 = 1$.  }
    \label{fig:gauss_ex_rtp}
\end{figure*}

The mean of the run-and-tumble particle takes a form similar to the passive Brownian case, with an additional contribution coming from the potential asymmetry in how the internal velocity states $\sigma(t)$ are reset:
\begin{equation}
    \langle x \rangle_* = \frac{r v_0 (\rho_+-\rho_-) }{(r + 2 \alpha  )(r + k \mu )}  + \frac{r}{r + k \mu } \langle x \rangle_R.
\end{equation}
When the internal state is reset asymmetrically $\rho_+ \neq \rho_-$, the mean obtains a more complex behavior. Indeed, even for $\langle x \rangle_R = 0$, the first term alone gives rise to non-monotonic behavior as a function of reset rate.  This is due to the competition between the spatial resetting  and the asymmetric velocity resets. In this case, the mean is maximized at $r = \sqrt{ 2 k \mu \alpha}$, in agreement with the numerics in Fig (\ref{fig:rtp_k1} B), where again a Gaussian resetting distribution was considered. Interestingly, this maximum of the mean is independent of the degree of asymmetry $|\rho_+-\rho_-|$ between the velocity states. Also, note that for symmetric velocity resets $\rho_\pm = 1/2$ it is not possible to distinguish an active RTP from a passive Brownian by looking solely at the mean.

The fluctuations in the steady state are characterized by the variance, which reads
\begin{align}
  \kappa_2 & =  \frac{2 v_0^2}{(r + 2 k\mu)( r + 2 \alpha + k\mu)}  \nonumber \\
   &+ \frac{2 r v_0 (\rho_+ - \rho_-) \langle x \rangle_R}{(r + 2 k\mu)( r + 2 \alpha + k\mu)} \nonumber \\
   &-  \frac{[r v_0(\rho_+-\rho_-) + r (r+2\alpha)\langle x \rangle_R]^2}{(r + 2\alpha)^2 (r + k\mu)^2}  \nonumber \\
   &+ \frac{r}{r+ 2 k\mu}\langle x^2 \rangle_R
\end{align}
Fig. (\ref{fig:gauss_ex_rtp}) shows the variance as a function of the resetting rate for Gaussian resetting distribution. As in the passive case, we see a non-monotonic behavior at sharp resetting distributions for $z\neq 0$, while large resetting widths $\sigma$ give rise to monotonic variance.  

In all calculations for the run-and-tumble particle,  the passive limit can be taken to ensure consistency with the Brownian results. This limit can be taken by letting $v_0 \to \infty$, $\alpha \to \infty$, while keeping the ratio equal to a constant proportional to the diffusion coefficient. This reproduced the results of the previous sections.

\section{Conclusion}\label{sec:concl}
The steady-states under stochastic resetting to a distribution was studied analytically and the results supported by numerical simulations. We have proposed a method for calculating moments of any order in the steady state, with results valid for arbitrary resetting distributions $p_R(x)$.   We show that for homogeneous systems with translational invariance, distributed resetting will not introduce new phenomena, while for non-homogeneous ones the presence of multiple scales may give rise to novel behavior. This includes a strongly non-monotonic dependence of the cumulants on resetting rate in some cases, an effect which can be washed away by introducing too broad resetting distributions.

The presented results are useful when exact calculation of the steady state probability density is not possible due to potentially complex nature of the resetting distribution. Distributed resetting positions also closely mimics resetting schemes used in experimental setups with optical tweezers. We exemplified the framework by considering both passive Brownian particles and active run-and-tumble particles confined by harmonic potentials in one dimension. In both cases we calculate exact expressions for the moments, which we further study for a Gaussian resetting distribution. Numerical simulations show excellent agreement with the theoretical predictions.

In future works, it would be interesting to extend Eq. (\ref{eq:genform}) and Eq. (\ref{eq:coefs}) to other resetting schemes, including non-Poissonian waiting time distributions. It would also be interesting to compare the presented predictions to experiments on resetting colloids using optical tweezers, where resetting distributions naturally appear.

\begin{acknowledgements}
The author would like to thank Supriya Krishnamurthy, Francesco Mori, Francesco Coghi and Prabal S. Negi for insightful discussions and interactions. KSO acknowledges the Nordita fellowship program. Nordita is partially supported by Nordforsk. KSO acknowledges support by the Deutsche Forschungsgemeinschaft (DFG) within the project LO 418/29-1. 
\end{acknowledgements}

%\appendix

%\bibliographystyle{apsrev4-2}
%\bibliography{OlsenEtAl.bib}

%embedded bibliography


\end{document}